\documentclass[aps, prb, twocolumn, 10pt, superscriptaddress,longbibliography]{revtex4-2}
\usepackage{amsmath}
\usepackage{amsfonts}
\usepackage{amssymb}
\usepackage{mathtools}
\usepackage{bm}
\usepackage{cases}
\usepackage{braket}
\usepackage[version=4]{mhchem}
\usepackage{graphicx}
\usepackage{hyperref}
\hypersetup{
    colorlinks = true,
    citecolor = blue,
    pdfborder = 0 0 0
}
\usepackage{booktabs}
\allowdisplaybreaks%

\begin{document}
\title{Topological chiral spin liquids and competing states
\\
in triangular lattice SU($N$) Mott insulators}
\author{Xu-Ping Yao}
\affiliation{Department of Physics and HKU-UCAS Joint Institute 
for Theoretical and Computational Physics at Hong Kong, 
The University of Hong Kong, Pokfulam Road, Hong Kong, China}
\author{Yonghao Gao}
\affiliation{State Key Laboratory of Surface Physics and Department of Physics, 
Fudan University, Shanghai 200433, China}
\author{Gang Chen}
\email{gangchen@hku.hk}
\affiliation{Department of Physics and HKU-UCAS Joint Institute 
for Theoretical and Computational Physics at Hong Kong, 
The University of Hong Kong, Pokfulam Road, Hong Kong, China}

\date{\today}

\begin{abstract}
SU($N$) Mott insulators have been proposed and/or realized in    
solid-state materials and with ultracold atoms on optical lattices.
We study the two-dimensional SU($N$) antiferromagnets on 
the triangular lattice. Starting from an SU($N$) Heisenberg model
with the fundamental representation on each site in the large-$N$ limit, 
we perform a self-consistent calculation and find a variety of ground   
states including the valence cluster states, stripe ordered states 
with a doubled unit-cell and topological chiral spin liquids. The 
system favors a cluster or ordered ground state when the number 
of flavors $N$ is less than 6. It is shown that, increasing the number   
of flavors enhances quantum fluctuations and eventually transfer 
the clusterized ground states into a topological chiral spin liquids. 
This chiral spin liquid ground state has an equivalent for the square 
lattice SU($N$) magnets. We further identify the corresponding 
lowest competing states that represent another distinct type of chiral 
spin liquid states. We conclude with a discussion about the relevant
systems and the experimental probes.
\end{abstract}

\maketitle

\section{Introduction}
\label{sec:I}

SU($N$) Mott insulators are representative examples of quantum systems 
with a large local Hilbert space where quantum fluctuations can be 
strongly enhanced and exotic quantum phases could be stabilized. 
All the SU($N$) spin operators are present in the effective model 
for the Mott insulators and would be able to shuffle all the spin states 
rather actively in the local Hilbert space. The system can be quite 
delocalized within the local Hilbert space, that is to enhance 
quantum fluctuations and induce exotic quantum phases. This aspect 
is fundamentally different from the SU(2) Mott insulators with 
large-$S$ local moments that also has a large local Hilbert space. 
For the large-$S$ SU(2) Mott insulators, the pairwise Heisenberg interaction
is quite ineffective to delocalize the spin states in the large-$S$ 
Hilbert space, and thus quantum fluctuations are strongly suppressed. 
Thus, it is the conventional wisdom not to search for exotic quantum phases
among the large-$S$ SU(2) Mott insulators, but among the spin-1/2 quantum 
magnets with a strong frustration. In contrast, the emergence of the SU($N$) 
Mott insulators brings a new searching direction for exotic quantum phases. 

SU($N$) Mott insulators are not a theoretical fantasy, but exist in nature.  
It has been shown that the ultracold alkaline-earth atoms (AEA) on optical lattices 
can simulate quantum many-body physics with an SU($N$) symmetry 
without any fine-tuning~\cite{Gorshkov2010}. 
The nuclear spin of fermionic AEA can be as large as ${I=9/2}$ for \ce{^87Sr}, while the outer shell 
electrons give a total spin ${S=0}$ and makes the hyperfine coupling inactive. 
This observation effectively extends the realization of SU($N$) 
magnets up to ${N=2I+1}$. 
Early efforts by Congjun Wu explored total spin-3/2 alkaline fermions on optical 
lattices where the SU(4) symmetry can be achieved with fine 
tuning~\cite{PhysRevLett.91.186402,PhysRevB.77.134449,PhysRevLett.95.266404,WU_2006}. 
Quantum Monte Carlo simulations were introduced later to study magnetic properties of the SU$(2N)$ Hubbard model~\cite{PhysRevB.88.125108,PhysRevLett.112.156403}. 
There is also some effort in searching for an emergent SU($N$) 
symmetry in real materials particularly for the SU(4) case. 
The two-orbital Kugel-Khomskii model can become SU(4) symmetric
after some fine tuning~\cite{Kugel__1982,PhysRevB.60.6584}. 
Experimental and numerical evidence also suggests that \ce{Ba3CuSb2O9} could be a prominent candidate on a decorated honeycomb lattice~\cite{PhysRevB.100.205131}, though the Cu-Sb dumbbell is quenched rather than an active degree of freedom.  
The SU(4) Heisenberg model has further been proposed for the spin-orbit-coupled
Mott insulator $\alpha$-\ce{ZrCl3} where the degree of freedom is the spin-orbit-entangled
${J=3/2}$ local moment~\cite{PhysRevB.82.174440} on a honeycomb lattice~\cite{PhysRevLett.121.097201}. 
\begin{figure}[b]
    \centering
    \includegraphics[width=0.48\textwidth]{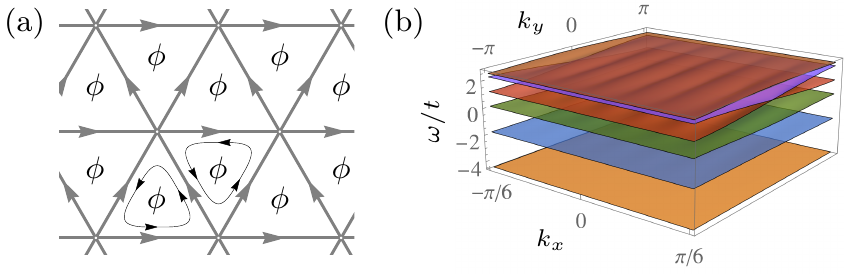}
    \caption{(a) Triangular lattice with background U(1) gauge flux $\phi$ within each plaquette. 
    (b) The spinon dispersions for chiral spin liquids with $\phi=5\pi/6$.}
    \label{fig:01}
\end{figure}
More recently, it has been proposed that the Mott insulating and 
superconducting behaviors in twisted bilayer graphenes can be captured by a two-orbital 
Hubbard model with an emergent SU(4) symmetry on a moir\'{e} triangular 
lattice~\cite{PhysRevLett.121.087001}. Other work even suggested that double moir\'{e} layers, 
built from transition metal dichalcogenides or graphenes, separated from one another 
by a thin insulating layer would be a natural platform to realize Hubbard models on the 
triangular lattice with SU(4) or SU(8) symmetries~\cite{Ashvin:2020,zhang2021su4}.


Owing to enhanced quantum fluctuations for SU($N$) Mott insulators,
the pioneering theoretical works~\cite{PhysRevLett.103.135301,PhysRevB.84.174441} 
by Hermele \textit{et al}, have obtained 
the topological chiral spin liquid (CSL) ground states with 
intrinsic topological orders even for an unfrustrated square lattice 
when ${N \geq 5}$~\cite{PhysRevLett.103.135301,PhysRevB.84.174441,PhysRevA.93.061601}.  
Depending on the atom occupation numbers, the system can support both Abelian 
and non-Abelian statistics for the anyonic excitations. Since unfrustrated 
lattices such as square and honeycomb lattices already bring exotic and interesting 
physics~\cite{PhysRevLett.107.215301,PhysRevLett.105.265301,PhysRevX.2.041013}, 
the frustrated lattices could further harbor nontrivial quantum phases, for example the triangular lattice with SU(3) and SU(4) spins \cite{10.21468/SciPostPhys.8.5.076,PhysRevB.101.245159,PhysRevResearch.2.023098}. 
In this work, we focus on the SU($N$) Heisenberg model
on a triangular lattice where each lattice site comprises the fundamental 
representation of the SU($N$) group. For AEA, 
this corresponds to one atom per lattice site and is known to be most stable
against the three-body loss. Because the $1/N$ filling is kept throughout, 
the large-$N$ limit differs fundamentally from the large-$N$ extension 
of the spin-1/2 SU(2) models where the 1/2 is kept and two sites can form 
a SU(2) spin singlet~\cite{Subir}. Here, as $N$ sites are needed to form a singlet, the 
valence cluster solid (VCS) state is generically disfavored in the large-$N$ limit. 
Instead, by our large-$N$ calculation, two types of CSL states with background U(1) gauge flux $\phi = \pi-\pi/N$ and $\pi-\pi/(2N)$ (see Fig.~\ref{fig:01}(a)) are identified as the ground and lowest competing states for $6 \le N \le 9$, respectively. 
For smaller $N$'s, various symmery-broken cluster/stripe states are obtained. 
We expect the large-$N$ results are more reliable 
when $N$ is large. 

\begin{table}[t]
    \caption{
    The SCM results of the ground and 
    lowest competing states for ${2 \le N \le 9}$. The average background fluxes of doubled unit-cell stripe states for $N=4$ and $5$ satisfy $\phi_{\text{avg}}=\pi-\pi/N$.}
    \label{tab:A} 
    \begin{ruledtabular}
        \begin{tabular}{ccc}
            $N$  & Ground state & Lowest competing state \\
            \midrule
            2 & dimer state & CSL $\phi = \pi/2$ \\
            3 & three-site complex VCS & CSL $\phi = 2\pi/3$ \\
            4 & four-site VCS &  doubled unit-cell stripe\\
            5 & doubled unit-cell stripe & CSL  $\phi = 4\pi/5$\\
            $6\le N \le 9$ & CSL $\phi = \pi-\pi/N$ & CSL $\phi = \pi-\pi/(2N)$
        \end{tabular}
        \end{ruledtabular}
\end{table}

The rest of this paper is organized as follows. 
The general Heisenberg model of SU($N$) spins is introduced and simplified at the large-$N$ saddle point in Sec.~\ref{sec:II}.
The self-consistent minimization algorithm is implemented to solve the reduced mean-field spinon Hamiltonian.
The technique details of this algorithm are described in Sec.~\ref{sec:III}. 
It is emphasized that the optimized solutions strictly satisfy the local constraints. 
In Sec.~\ref{sec:IV}, both the ground states and lowest competing states for $2 \le N \le 9$ are reported and analysed, especially for two types of CSL states. 
Finally, in Sec.~\ref{sec:V}, this work conclude with a discussion about relevant systems and the experimental probes.

\section{The SU($N$) Heisenberg model in the large-$N$ approximation}
\label{sec:II}

We begin with an SU($N$) Heisenberg model 
on the triangular lattice where each site comprises 
the fundamental representation of the SU($N$) group. 
This model can be obtained from the strong coupling limit of an
SU($N$) Hubbard model with $1/N$ filling or one particle per site.
The SU($N$) Heisenberg model is given as 
\begin{equation}
\label{eq:SpinHam}
    {\mathcal H}=J \sum_{\langle {\boldsymbol r} {\boldsymbol r}' \rangle} 
    S_{\alpha\beta}({\boldsymbol r})S_{\beta\alpha}({\boldsymbol r}'),
\end{equation}
where $J$ is the antiferromagnetic exchange interaction and the sum is 
taken over the nearest neighbor bonds and spin flavors.
The SU($N$) spin operators 
can be simply expressed with the Abrikosov fermion representation
${S_{\alpha\beta}({\boldsymbol r})=f_{{\boldsymbol r}\alpha}^{\dagger}
f_{{\boldsymbol r}\beta} }$, and ${\alpha,\beta=1,\ldots, N}$. 
Hereafther, a summation over repeated indices in the form of Greek letters is supposed unless otherwise specified. 
A local constraint on the fermion occupation  
${f_{{\boldsymbol r}\alpha}^{\dagger}f_{{\boldsymbol r}\alpha}=1}$
is imposed to reduce the enlarged Hilbert space.
In principle, an SU$(N$) singlet could be formed by $N$ sites. But the SU($N$) exchange rapidly transforms one $N$-site singlet to other sets of $N$-site singlets, resulting in a failure of the conventional understanding.
Instead, the spin Hamiltonian 
Eq.~\eqref{eq:SpinHam} is solvable in the limit $N \rightarrow \infty$ 
via a large-$N$ saddle point approximation in the imaginary-time
path integral formulation. 
The corresponding partition function can be expressed as 
\begin{eqnarray}
{\mathcal Z} = \int  {\mathcal D} \chi^{\dagger}  {\mathcal D} \chi {\mathcal D} \mu 
{\mathcal D} f^{\dagger} {\mathcal D} f e^{- {\mathcal{S}}}.
\end{eqnarray}
The action ${\mathcal S}$ is given as 
\begin{widetext}
\begin{eqnarray}
{\mathcal S} &=& \int_0^{\beta} d\tau \Big[
\sum_{\boldsymbol r} f^{\dagger}_{{\boldsymbol r}\alpha}  \partial_\tau^{}  f^{}_{{\boldsymbol r}\alpha}
+ \sum_{\langle  {\boldsymbol r} {\boldsymbol r}'\rangle } 
\big(\chi_{  {\boldsymbol r} {\boldsymbol r}' }^{}
f^{\dagger}_{{\boldsymbol r}\alpha} f^{}_{{\boldsymbol r}'\alpha} + \mathrm{ h.c.}
\big)
+ N \sum_{\langle  {\boldsymbol r} {\boldsymbol r}'\rangle }
\frac{|\chi_{ {\boldsymbol r}{\boldsymbol r}'}^{}|^2}{\mathcal J} + \sum_{\boldsymbol r} \mu_{\boldsymbol r}^{}
( f^\dagger_{{\boldsymbol r}\alpha} f^{}_{\boldsymbol r \alpha}- 1) 
\Big],
\end{eqnarray}
where $\mu_{\boldsymbol r}$ is the Lagrangian multiplier to enforce the Hilbert 
space constraint, $\chi_{ {\boldsymbol r}{\boldsymbol r}'}$ is the auxiliary field
to decouple the fermion operators, and ${{\mathcal J} \equiv NJ}$. As the action 
${\mathcal S}$ scales linearly with $N$, the large $N$ limit leads to a saddle point 
approximation that results in the saddle point equations, 
${\chi_{ {\boldsymbol r}{\boldsymbol r}'}^{}=-{\mathcal{J}}\langle {f_{ {\boldsymbol r}'\alpha}^{\dagger} f^{}_{{\boldsymbol r}\alpha}} \rangle/N}$, ${\langle {f_{ {\boldsymbol r}\alpha}^{\dagger} f_{ {\boldsymbol r}\alpha}^{}} \rangle =1}$, and
the saddle point or mean-field Hamiltonian for the fermionic spinons is
\begin{eqnarray}
\label{eq:MFTHam}
    \mathcal{H}_{\text{MF}}& =& 
    \frac{N}{\mathcal{J}}\sum_{\langle {\boldsymbol{rr'}} \rangle}
    |\chi_{\boldsymbol{rr}'}^{}|^2
    +\sum_{\langle \boldsymbol{rr}' \rangle}
     (\chi_{\boldsymbol{rr}'}^{}f_{\boldsymbol{r}\alpha}^{\dagger}f_{\boldsymbol{r}'\alpha}^{}+ \mathrm{h.c.})
+\sum_{\boldsymbol{r}}\mu_{\boldsymbol{r}}^{}(1-f_{\boldsymbol{r}\alpha}^{\dagger}
f^{}_{\boldsymbol{r}\alpha}).
\end{eqnarray} 
\end{widetext}
In the following, we will search for the saddle point with the 
lowest mean-field energy $E_{\text{MF}}$ numerically and discuss the properties 
of ground states.  
Before that, we first discuss the lower bound of 
$E_{\text{MF}}$ and the bound saturation conditions for reference. 
An exact lower bound on $E_{\text{MF}}$ 
for generic lattices~\cite{PhysRevB.42.2526} was first obtained by 
Rokhsar for the half-filling, and was shown to  
be saturated by valence bond states with various spin singlet coverings. 
These states break the lattice translation but preserve 
the spin-rotation symmetry, and fluctuations beyond the mean field 
can break the high degeneracy among them~\cite{Subir}. 
The lower bound was generalized to 
the $1/N$ filling with~\cite{PhysRevB.84.174441}
\begin{equation}
\label{eq:simplexbound}
    E_{\text{MF}} \ge - N_s \frac{N-1}{2 N}\mathcal{J}_{\text{max}},
\end{equation}
where $N_s$ is the number of lattice sites. 

We set ${\mathcal{J}_{\text{max}}=\mathcal{J}}$ 
for each bond. The saturation for Eq.~\eqref{eq:simplexbound} 
is reached by a $N$-simplex VCS state composed by $N$-site simplices. 
That is, every site on the lattice is directly connected to the other ${N-1}$ sites 
within the same cluster by a single bond with an exchange coupling 
$\mathcal{J}_{\text{max}}$. 
On a $d$-dimensional lattice, the $N$-simplex VCS with $N>d+1$ is prohibited without fine-tuning of the exchange.
Thus, there are only two-simplex 
and three-simplex VCS's on the triangular lattice. For ${N \ge 4}$, 
possible cluster states are general $N$-site ones satisfying a stricter 
energy bound~\cite{PhysRevB.84.174441}
$E_{\text{MF}} \ge -{ \mathcal{N}_b\mathcal{J}_{\text{max}}}/{N}$,
where $\mathcal{N}_b$ is the total number of isolated bonds in the lattice.

\begin{table}[b]
    \caption{\label{tab:B} 
    The SCM results of the ground state and the lowest competing state energies for 
    ${2 \le N \le 9}$. 
    The energy is in units of $N\mathcal{J}N_s=N^2J N_s$.}
    \begin{ruledtabular}
        \begin{tabular}{c c c}
        $N$ & Ground state & Lowest competing state \\
        \midrule
        2 & $-0.1250000$ & $-0.1202034$ \\
        3 & $-0.1111111$ & $-0.0999171$ \\
        4 & $-0.0781250$ & $-0.0760440$ \\
        5 & $-0.0581877$ & $-0.0581046$ \\
        6 & $-0.0455285$ & $-0.0443810$ \\
        7 & $-0.0364651$ & $-0.0357994$ \\
        8 & $-0.0297864$ & $-0.0293744$ \\
        9 & $-0.0247509$ & $-0.0244826$ \\
        \end{tabular}
        \end{ruledtabular}
\end{table}

\section{The SCM algorithm}
\label{sec:III}

To determine the saddle-point solutions, we closely follow the numerical self-consistent minimization 
(SCM) algorithm developed in Refs.~\onlinecite{PhysRevLett.103.135301,PhysRevB.84.174441}. 
The technical details of the algorithm are described in the following. 
During one energy optimization process for a given geometry and periodic boundary conditions, the algorithm starts from initializing the fields $\chi_{\boldsymbol{rr}'}$ 
at each bond randomly with ${\chi_{\boldsymbol{rr'}}=|\chi_{\boldsymbol{rr}'}|e^{i\phi_{\boldsymbol{rr}'}}}$, 
with a uniform distribution of amplitudes ${|\chi_{\boldsymbol{rr}'}| \in [0.02,0.20]}$ and 
phases ${\phi_{\boldsymbol{rr}'}\in[0,2\pi]}$. 
The chemical potentials are set to be the default value $\mu_{\bm{r}}=0$ in the beginning. Obviously, the local constraints are violated here in general and the deviation of the local fermion density can be denoted as $\delta n_{\bm{r}} = 1 - \braket{f_{\bm{r}\alpha}^{\dagger}f_{\bm{r}\alpha}}$. The expectation value is taken using the ground state of $\mathcal{H}_{\text{MF}}$ with the unchanged $\mu_{\bm{r}}$ at this stage. To obtain the correct density, one must adjust the chemical potential $\mu_{\bm{r}}$ by $\delta \mu_{\bm{r}}$ at each site. It is found that to the lowest order, the desired adjustment of chemical potentials $\delta\mu_{\bm{r}}$ can be expressed as 
\begin{equation}\label{eq:response}
    \delta \mu_{\bm{r}} = \sum_{\bm{r}'}G^{-1}(\bm{r}-\bm{r}',0)\delta n_{\bm{r}'},
\end{equation}
where $G^{-1}(\bm{r}-\bm{r}',0)$ is nothing but the inverse of the density-density correlation at zero frequency. 
The elements of the correlation $G(\bm{r}-\bm{r}',0)$ can be calculated at the single particle level from the mean-field Hamiltonian $\mathcal{H}_{\text{MF}}$, with the help of the standard linear response theory in principle. 
However, it follows that the inversion of $G(\bm{r}-\bm{r}')$ naively diverges. 
Physically, this is because a uniform adjustment of the chemical potential at every site is trivial and the fermion density will keep intact. 
This obstacle is overcome by following the treatment suggested in Ref.~\onlinecite{PhysRevB.84.174441}; that is diagonalizing $G(\bm{r}-\bm{r}',0)$ and only focusing on its non-zero eigenvalues $g_{i}$. 
Note that the index $i$ refers to the basis where $G(\bm{r}-\bm{r}',0)$ is diagonalized. 
In such a basis, the relationship Eq.~\eqref{eq:response} becomes
\begin{equation}
    \delta \mu_{i} = \delta n_{i} / g_{i}.
\end{equation}
Here, the index $i$ should not be summed. 
For all indices with vanishing eigenvalues $g_i$, $\mu_{i}$ is taken to be zero concurrently. 
Then, the adjustment of the chemical potential $\delta \mu_{\bm{r}}$ is well-defined. 
A simple replacement of $\mu_{\bm{r}}\rightarrow\mu_{\bm{r}}+\delta \mu_{\bm{r}}$ gives a new mean-field Hamiltonian $\mathcal{H}_{\text{MF}}$ and related ground state. 
In consequence, it affects the local fermion density conversely, resulting in a new deviation $\delta n_{\bm{r}}$.
These processes construct a self-consistent procedure and the problem of searching for an appropriate set of chemical potential deviation $\delta \mu_{\bm{r}}$ can be solved by iterating the procedure until the density is uniform.
This is the core of the algorithm to preserve the local single-occupation constraints $n_{\bm{r}}=\braket{f_{\bm{r}\alpha}^{\dagger}f_{\bm{r}\alpha}}=1$.

\begin{figure}[t]
    \centering
    \includegraphics[width=0.48\textwidth]{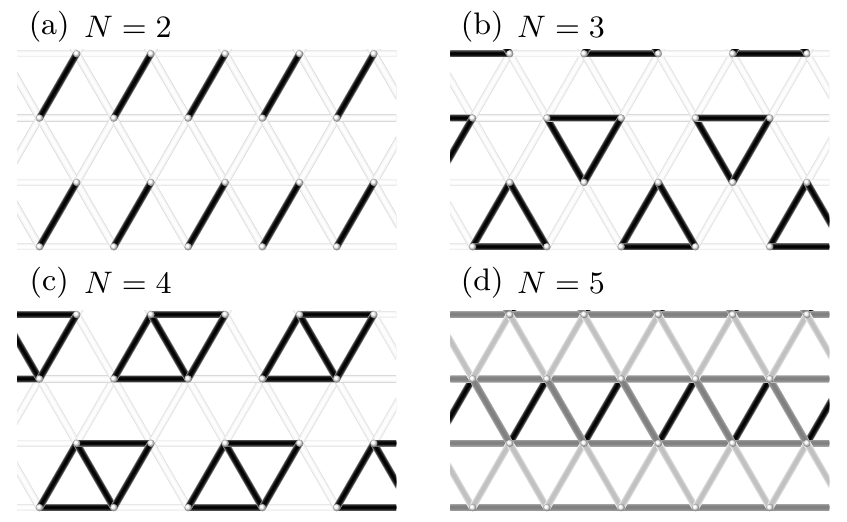}
    \caption{Ground states in the large-$N$ limit for (a) ${N=2}$, (b) ${N=3}$, (c) ${N=4}$ and (d) ${N=5}$. 
    For $2 \le N \le 4$, any state covering all sites by the same dimer/cluster unit has the same energy in each case. 
    We merely show one of such configurations. For ${N=5}$, there is a stripe pattern with doubled 
    unit-cell along one of lattice vectors. The grayness of bonds indicates the relative magnitude of expectation values $|\chi_{\boldsymbol{rr}'}|$. Black (white) bonds refer to maximal (zero) $|\chi_{\boldsymbol{rr}'}|$.}
    \label{fig:02}
\end{figure}

With the modified chemical potentials calculated in the previous step, an update set of auxiliary fields $\chi_{\bm{rr}'}$ can be determined via ${\chi_{\boldsymbol{rr}'}'=-{\mathcal{J}}\langle {f_{\boldsymbol{r}'\alpha}^{\dagger}f_{\boldsymbol{r}\alpha}} \rangle/N}$, but the local constraints may be violated once again. 
The amended auxiliary fields and chemical potentials can be treated as a new and better starting point. 
The procedure described in the previous step is thus implemented iteratively until reaching a converge of the ground state energy within a given numerical error. 
It has been shown in Ref.~\onlinecite{PhysRevB.84.174441} that the energy of the final state must be less or equal to that of the initial state after the optimization process. 
Therefore, we eventually obtain a local energy minimum. 
In order to reach the global minimum as much as possible, we start from at least 50 randomly initialized fields, and reap a collection of local minima satisfying the single-occupation constraints. 
The lowest two are accepted as the best results of the ground state and lowest competing state energies.

\section{The mean-field results}
\label{sec:IV}

We describe the ground states and lowest competing states of the mean-field Hamiltonian 
Eq.~\eqref{eq:MFTHam} from the SCM algorithm. Because different geometries can 
accommodate different candidate ground states especially cluster states with unknown dimensions, 
in the calculation, we consider all unit cells of a parallelogram geometry $\ell_1 \times \ell_2$ with 
${\ell_{1,2} \le 2N}$ for ${2 \le N \le 5}$ and ${\ell_{1,2} \le N}$ for ${N \ge 6}$, respectively. 
In some ordered cases for $N=4,5$, larger unit-cell sizes are also 
considered for confirmation. Each unit cell is repeated by $L_{1,2}$ $({\ge 20})$ times 
along the triangular Bravais lattice vectors to form a superlattice.
The superlattice itself has periodic boundary conditions. 
In practice, for each case, we ran the SCM procedure at least 50 times with different random seeds to avoid any missing of ground states caused by numerical problem. The results are listed in Table~\ref{tab:A} and Table~\ref{tab:B}.

\begin{figure}[t]
    \centering
    \includegraphics[]{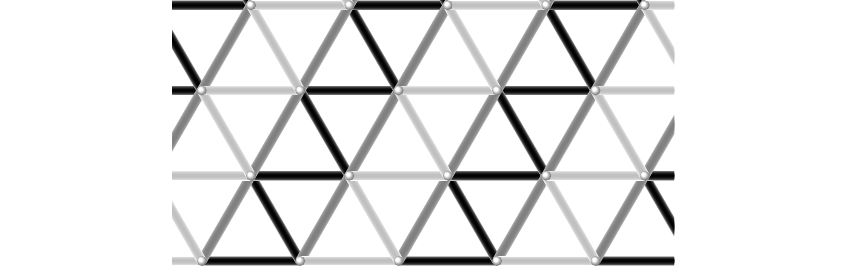}
    \caption{The lowest competing state in the large-$N$ limit for ${N=4}$. 
    This is a stripe pattern with doubled unit-cell along one of lattice vectors. 
    The grayness of bonds indicates the relative magnitude of expectation value $|\chi_{\boldsymbol{rr}'}|$.}
    \label{fig:03}
\end{figure}

For $N=2$ and $3$, the lowest energies we found saturate the bound Eq.~\eqref{eq:simplexbound}, meaning that the ground states are dimer and three-site simplex VCS, respectively. 
Actually, the obtained ground states are highly degenerate because any 
state with each site covered by one dimer/cluster unit has the 
same energy in the large-$N$ limit. An ordered dimer or three-site simplex VCS state, as illustrated in Figs.~\ref{fig:02}(a-b), is expected to be selected beyond the mean field.
Note that, the non-zero expectation values $|\chi_{\bm{rr}'}|$ in Fig.~\ref{fig:01}(a) and (b) can only take $1/2$ and $1/3$, respectively. 
The true ground state for $N=2$ is the well-known 
$120^{\circ}$ N\'{e}el state and differs from the mean-field here. 
The $N=3$ case was also shown to have a three-sublattice magnetic order in previous numerics
~\cite{PhysRevLett.97.087205,PhysRevB.85.125116}.
This reflects the deficiency of the 
large-$N$ approximation for small $N$ cases. In Table~\ref{tab:A}, we further find that 
the lowest competing states are CSL states with $\phi = \pi-\pi/N$.

The $N=4$ case is more compelling due to the rapid growth of experimental proposals \cite{PhysRevLett.117.167202,PhysRevLett.121.225303,Ashvin:2020,zhang2021su4} 
and large-$N$ approximation could provide useful insight here. We find the ground state 
energy do not saturate the bound Eq.~\eqref{eq:simplexbound} as excepted, but saturate 
the stricter bound discussed previously. The ground states are four-site VCS depicted in 
Fig.~\ref{fig:02}(c), accompanied by a large degeneracy for 
the same reason as $N=2$ and $3$. 
The non-zero expectation values are $|\chi_{\bm{rr}'}|=1/4$. 
A similar plaquette order was also reported in a recent work~\cite{zhang2021su4}. 
In Fig.~\ref{fig:03}, we depict the lowest competing state for ${N=4}$. 
One can see that the lattice is covered by stripes with three different bond expectation values $|\chi_{\boldsymbol{rr}'}|\approx 0.030$, $0.158$, and $0.224$. 
Along one of lattice vectors, there is a unit-cell doubling
that breaks the lattice translation. 
The background fluxes through each plaquette are inhomogeneous as well and manifest the same unit-cell doubling pattern. 
Specifically, the average flux is a constant value $\phi_{\text{avg}}=\pi-\pi/N$ where $N=4$.
The same ordered pattern has also been obtained through the DMRG study from the Kugel-Khomskii 
model~\cite{Chaoming:2019}. They attribute the symmetry breaking to symmetry allowed 
Umklapp interactions in certain finite geometries. In our results, however, such a stripy state is not the ground state even in the two-dimensional limit.

From ${N=5}$, neither the bound Eq.~\eqref{eq:simplexbound} nor 
the stricter one can be saturated. Thus, the ground states are no longer VCS. 
Our numerical calculation finds that the ground state for ${N=5}$ is very similar 
to the lowest competing state for $N=4$ (see Fig.~\ref{fig:02}(d)), except that the bond expectation values can only take $|\chi_{\boldsymbol{rr}'}|\approx 0.072$, $0.145$, and $0.179$, and the average background flux $\phi_{\text{avg}}=\pi-\pi/N$ shifts accordingly. 
It also breaks the lattice translation symmetry along one of lattice vectors 
and manifests itself as a stripe pattern with a doubled unit-cell. 
The lowest competing state for $N=5$ is a CSL with $\phi = 4\pi/5$. 
With further increasing $N$, the frustration is enhanced. Eventually, the ground states 
become CSL states with $\phi = \pi-\pi/N$ when ${6\le N \le 9}$. 
Correspondingly, the lowest competing states we found share the identical form. 
They are CSL states as well except that the background magnetic fluxes shift 
to $\phi = \pi -\pi/(2N)$.

\begin{figure}[t]
    \centering
    \includegraphics[]{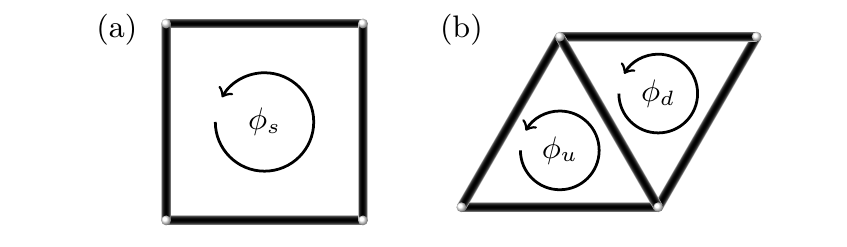}
    \caption{The definition of the emergent U(1) gauge flux $\phi$ for CSL on the (a) square and (b) triangular lattices. 
    The subscripts $s$, $u$ and $d$ refer to the square, up triangular and down triangular plaquettes, respectively.}
    \label{fig:04}
\end{figure}

With the two types of CSL states identified, we now discuss the properties of these topological liquid states. The CSL is characterized by the mean-field saddle point
\begin{gather}
    \chi_{\boldsymbol{rr}'}=|\chi| e^{i a_{\boldsymbol{rr}'}},\\
    \mu_{\boldsymbol{r}}=0.
\end{gather}
All bonds on the lattice have a uniform expectation value for $|\chi|$ but modulated by a 
U(1) gauge field $a_{\boldsymbol{rr}'}$ so that the flux $\phi$ on each plaquette is a constant. 
The CSL breaks both parity and time-reversal. The bond phase field $a_{\boldsymbol{rr}'}$ is 
treated as a fluctuating $\mathrm{U}(1)$ gauge field coupled to the fractionalized spinons. 
By checking the behavior of the mean-field Hamiltonian Eq.~\eqref{eq:MFTHam} at the CSL saddle points, 
we find that both types of CSL states have a fermion band structure with $N$ bands where 
only the lowest is filled (see Fig.~\ref{fig:01}(b)). The Fermi level lies in the gap between the lowest two bands, 
thus all discussion can be applied to both CSL states. 
Furthermore, the first type of CSL with $\phi_{u,d}=\pi-\pi/N$ on the triangular lattice can be mapped to the counterpart on a square lattice up to a time-reversal. If we regard the adjacent up and down triangles as a unit shown in Fig.~\ref{fig:04}, the phase of the shared bond has no contribution to the total flux, and we have the relationship
\begin{equation}
    \phi_{u}+\phi_{d}=-\phi_{s} \mod 2\pi,
\end{equation}
where $\phi_s$ is the background U(1) flux through each square plaquette. In Ref.~\cite{PhysRevLett.103.135301}, Hermele \textit{et al}, found that, the CSL states are ground states for $5 \le N \le 10$ on the square lattice in the mean-field calculation.
As the spinon is gapped out by the emergent U(1) gauge flux pattern, 
the Chern-Simons term enters the theory for U(1) gauge fluctuations. 
After integrating the gapped spinons, one obtains a topological quantum field theory
with a Chern-Simons term corresponding to the chiral Abelian topological order 
and anyonic statistics. The spinon is converted in anyons with a statistical angle ${\pi \pm \pi/N}$. 
Gapless chiral states carrying spin degree of freedom are also supported by the CSL as edge modes and the low-energy theory is described by the SU(N)$_1$ WZW model.

\section{Discussion}
\label{sec:V}

To summarize, we study the Heisenberg antiferromagnets with SU($N$) symmetry on the triangular lattice. 
In the large-$N$ approximation, a variety of ground states and lowest competing states are identified for different parameter $N$.
At the mean-field level, the ground state for $2 \le N \le 4$ is a $N$-site VCS state 
with a large degeneracy. Specifically, ordering patterns with doubled unit cell and average background flux $\phi_{\text{avg}}=\pi-\pi/N$ are found 
in the $N=4$ and $5$ cases. These ordered states break the lattice translation symmetry 
along one of lattice vectors and become the lowest competing state and ground state for 
$N=4$ and $5$, respectively. The frustration from SU($N$) exchange interaction 
is enhanced when $N > 5$, resulting two types of CSL states as the lowest two states for 
$6 \le N \le 9$. Among them, the CSL states with $\phi=\pi-\pi/N$ have a lower energy, 
and have its counterpart on the square lattice. Although the true ground states 
are not what we found for $N<4$, our calculation can provide useful insight for 
cases $N \ge 4$ where large-$N$ approximation becomes more reliable.
In a very recent DMRG study of an SU(4) spin model on the triangular lattice, phase diagrams for integer fillings were obtained and compared with conventional MFT ones \cite{zhang2021su4}. 
The quite good agreement of the phase boundaries determeted by two methods suggests that $N = 4$ is perhaps large enough for the mean-field analysis we performed in this work.

Thanks to the development of ultracold experimental techniques, 
the SU($N$) Mott insulators have been realized with AEA 
on various optical lattices using the Pomeranchuk cooling~\cite{Taie_2012},
even the Mott crossover and SU($N$) antiferromagnetic spin correlations were recently observed
with \ce{^173Yb} atoms~\cite{PhysRevX.6.021030,taie2020observation}.
Nontrivial physics of multicomponent fermions with broken SU$(N)$ symmetry were also proposed on this platform~\cite{PhysRevA.98.063628}.
The emergency of the widely concerned SU(4) or even SU(8) symmetric
interaction has been proposed in the twisted bilayer graphene 
and double moir\'{e} layers systems recently~\cite{Ashvin:2020,PhysRevLett.121.097201}. 
This moir\'{e} lattice system could be an novel platform 
to detect possible phases in this work. 
The ultracold atom systems may have many restrictions 
in the detection of anyonic spinon excitations and edge states.
 Nevertheless, spin-dependent Bragg
spectroscopy may be used to detect the spinon continuum~\cite{PhysRevB.84.174441,
PhysRevA.93.061601,PhysRevLett.103.135301},
 singlet-triplet oscillation technique can detect the nearest-neighbor spin correlation~\cite{taie2020observation} for CSLs, 
 and the lattice potential could be adjusted to 
 localize and manipulate the anyonic quasiparticles~\cite{PhysRevA.93.061601,PhysRevLett.103.135301,PhysRevB.84.174441}.
 For solid-state systems, more techniques are available. 
 These include but are not restricted to quantized thermal Hall transport~\cite{Gao_2020}
 for the edge modes, scanning tunneling microscope of anyons at defects~\cite{PhysRevX.8.011037},
 or even angle-resolved photon emission measurement for the 
spinon signatures~\cite{PhysRevB.87.045119}.

\begin{acknowledgments}
We thank Shizhong Zhang for some discussion 
and the proofreading of the manuscript. This work is supported by the
research funds from the Ministry of Science and Technology of China 
with grant No.2016YFA0301001, No.2018YFE0103200 and No.2016YFA0300500, 
by Shanghai Municipal Science and Technology Major Project with Grant        
No.2019SHZDZX04, and from the Research Grants Council of Hong Kong 
with General Research Fund Grant No.17303819 and No.17306520.
\end{acknowledgments}

\bibliography{TriSUNRef.bib}

\end{document}